\documentclass[12pt]{article}
\usepackage[cp1251]{inputenc}
\usepackage[russian]{babel}
\usepackage{amsfonts}

\newtheorem{theorem}{Theorem}

\newtheorem{corollary}{Corollary}

\begin{document}

\title{Discrete analogues of the Dixmier operators}
\author{А.Е. Мironov}

\date{}

\maketitle

\section{Introduction and main results}
In this paper we construct  discrete analogues of the Dixmier
operators, that is, commuting difference operators corresponding
to a spectral curve of genus 1 whose coefficients are polynomials
of the discrete variable $n$.

The theories of commu\-ting ordinary differential operators and
commu\-ting difference operators have a lot in common, therefore,
to begin with, we recall some fundamental results concerning
commuting differential operators.

According to the Burchnall-Chaundy Lemma [1], if the two differential operators
$$
 L_1=\frac{d^n}{dx^n}+u_{n-1}(x)\frac{d^{n-1}}{dx^{n-1}}+\dots+u_0(x),\
 L_2=\frac{d^m}{dx^m}+v_{m-1}(x)\frac{d^{m-1}}{dx^{m-1}}+\dots+v_0(x)
$$
commute, then there exists a nonzero polynomial $Q(\lambda,\mu)$
of two commuting variables $\lambda,\mu$, such that
$$
 Q(L_1,L_2)=0.
$$
On the plane ${\mathbb C}^2$ with the coordinates $(\lambda,\mu)$
this polynomial defines the {\sl spectral curve}
$$
 \Gamma=\{ (\lambda,\mu)\in {\mathbb C}^2:Q(\lambda,\mu)=0\}.
$$
Suppose that $\Gamma$ be a smooth curve. Considering some common
eigen-function $\psi$ of the operators $L_1$ and $L_2$, one can
show that the corresponding eigen-values $\lambda$ and $\mu$ are
connected by a polynomial relation $Q(\lambda,\mu)=0$ (see [2]).
Thus, the spectral curve parametrizes the common eigen-functions
and eigenvalues of $L_1$ and $L_2$
$$
 L_1\psi(x,P)=\lambda\psi(x,P),\ \ L_2\psi(x,P)=\mu\psi(x,P), \ \
 P=(\lambda,\mu)\in\Gamma.
$$
 The {\sl rank} of operators $L_1,L_2$ is called dimension of space of
common eigen-functions with fixed eigenvalues
$(\lambda,\mu)\in\Gamma$.

In the case of operators of rank $l=1$, one can find the function
$\psi(x,P)$ (Baker-Akhiezer function) in explicit form by means of
the theta-function of the  Jacobi variety of the curve $\Gamma$,
and in this case the operators' coefficients are easily found.
Krichever [2] gives a classification of operators of rank $l>1$.
But the problem of finding the operators' coefficients is not
solved in the general case. The following particular results
relating to operators of rank more than 1 are known: Dixmier [3]
found examples for operators of rank 2 corresponding to an
elliptic curve with polynomial coefficients
$$
 L_1=\left(\frac{d^2}{dx^2}-x^3-\alpha\right)^2-2x,\
$$
$$
L_2=\left(\frac{d^2}{dx^2}-x^3-\alpha\right)^3-
\frac{3}{2}\left(x\left(\frac{d^2}{dx^2}-x^3-\alpha\right)+
 \left(\frac{d^2}{dx^2}-x^3-\alpha\right)x\right),
$$
where $\alpha$ is some constant. Krichever and Novikov [4] found
all operators of rank 2 corresponding to the elliptic curve, and
Grinevich [5] found spectral data corresponding to operators with
rational coefficients. Mokhov [6] found operators for $l=3$, also
corresponding to the elliptic curve. In [7] and [8], we found
examples for operators with rank $2$ corresponding to the curve of
genus $g=2$, amongst which there are operators with polynomial
coefficients, and also formally self-adjoined operators.

As for smooth operators, for commuting difference operators of the form
$$
 L_1=\sum^{N_+}_{N_-}u_i(n)T^i,\
 L_2=\sum^{M_+}_{M_-}v_i(n)T^i,
$$
where $n\in{\mathbb Z}$ is a discrete variable, and $T$ --- the
shift operator on the discrete variable
$$
 Tf(n)=f(n+1),
$$
there exists a spectral curve $\Gamma$ given in ${\mathbb C}^2$ by
a polynomial $Q(\lambda,\mu)$, parametri\-zing their common
eigen-functions and eigenvalues
$$
L_1\psi(n,P)=\lambda\psi(n,P),\ \ L_2\psi(n,P)=\mu\psi(n,P), \ \
P=(\lambda,\mu)\in\Gamma.
$$
Define the rank of operators $L_1$ and $L_2$ in the same way as
for the smooth case as a dimension of space of common
eigen-functions in the point in general position $P\in\Gamma$. One
of the basic differences between the discrete and the smooth case
consists of the following. Any commutative ring of ordinary
differential operators is isomorphic to a ring of meromorphic
functions on the algebraic curve with pole only in the point
$Q\in\Gamma$, whereas any commutative ring of difference operators
is isomorphic to a ring of meromor\-phic functions on the
algebraic curve with $m$ poles, where $m$ can be any natural
number [9]. Such operators are called $m$-point operators.

Mumford [10] and Krichever [11] found spectral data corresponding
to two-pointed operators of rank 1.

In this paper we will consider only single-pointed operators. For
$l>1$, finding the function $\psi(n,P)$ reduces to solving a
Riemann problem, and it can not be found in the explicit form.
 Krichever and Novikov [9] (see also [4]) suggest a method for
finding the operators' coefficients called method of Turin
parameters deformation, and this method does not require to find
$\psi(n,P)$. They show that the operators' coefficients can be
restored from solutions of equations on Turin parameters of
holomorphic stable vector bundles that are uniquely given by
function $\psi(n,P)$, with the operators' coefficients depending
on arbitrary $l$ of functional parameters. Namely, they show that
in order to restore the operators' coefficients it suffices to
find the matrix function
$$
 \chi(n,P)=\Psi(n+1,P)\Psi^{-1}(n,P),
$$
where $\Psi(n,P)$ is a Wronski matrix (see below) built on some
basis in the space of common eigen-functions. Applying this
method, Krichever and Novi\-kov found operators of rank 2
corresponding to the elliptic curve. Simultane\-ous\-ly, the
operators' coefficients are expressed by $\zeta$ and
$\wp$-Weierstrass functi\-ons of two functional parameters.

In this paper we show spectral data for operators of rank 2
corresponding to the elliptic curve and whose coefficients are
expressed by elementary functi\-ons of functional parameters. In
particular, we find operators that have polynomial coefficients,
just like the Dixmier operators.

We can assume the affine part of the spectral curve $\Gamma$ to be
given in
${\mathbb C}^2$ with the coordinates $(z,w)$ by equation
$$
 w^2=F(z)=z^4+c_2z^2+c_1z+1.
$$

Take $Q=(0,1)\in\Gamma$ as preferred point. $Q=(0,1)\in\Gamma$ (the
ring of commuting difference operators will be isomorphic to a ring of
meromorphic functions on $\Gamma$ with pole in $Q$).
The curve $\Gamma$ allows a holomorphic involution
$$
 \sigma:\Gamma\rightarrow\Gamma,\ \sigma(z,w)=(z,-w).
$$
At $l=2$ the matrice $\chi(n,P),\ P=(z,w)\in\Gamma$ has the form
$$
\chi(n,P)=
\left(
\begin{array}{cc}
 0 & 1 \\
 \chi_1(n,P) & \chi_2(n,P)
\end{array}\right).
$$
We consider the case where the involution $\sigma$ does
not change $\chi_1$, e.g.
$$
 \chi_1(n,P)=\chi_1(n,\sigma (P)),\ P\in\Gamma.\eqno{(1)}
$$
Then, the following holds:

\begin{theorem}
The functions $\chi_1(x,P)$ и $\chi_2(x,P)$ have the form
$$
\chi_1(n,P)=\frac{c(n)}{z-\gamma(n)}+\frac{c(n)}{\gamma(n)-\gamma(n+1)},
$$
$$
 \chi_2(n,P)=\frac{1}{2z}+\frac{a(n)}{2(z-\gamma(n))}+
\frac{w\gamma(n)}{2z(\gamma(n)-z)}+d(n),
$$
where
$$
c(n)=\frac{\gamma(n-1)(a^2(n)-F(\gamma(n))}{4\gamma(n)(\gamma(n)-\gamma(n-1))},\eqno{(2)}
$$
$$
d(n)=\frac{(a(n+1)-1)\gamma(n)+(a(n)+1)\gamma(n+1)}{2(\gamma(n)-\gamma(n+1))\gamma(n+1)},
$$
$\gamma(n),a(n)$ are arbitrary functions of the discrete variable
$n\in{\mathbb Z}.$
\end{theorem}

The function $\lambda_1(z)$ on the curve $\Gamma$, with the only
second-order pole in $Q$, looks as follows:
$$
 \lambda_1=\frac{1}{2z^2}+\frac{c_1}{4z}+\frac{w}{2z^2}.
$$
Let $b_i(n),e_i(n),p_i$ denote the series decomposition coefficients
of functions $\chi_1,$ $\chi_2,$ and $\lambda_1$ in the neighbourhood
of $Q$:
$$
\chi_1(n,z)=b_0(n)+b_1(n)z+\dots, \ \ \
\chi_2(n,z)=\frac{1}{z}+e_0(n)+e_1(n)z+\dots,
$$
$$
\lambda_1=\frac{1}{z^2}+\frac{p_1}{z}+p_0+\dots, \ \ \
p_0=-\frac{c_1^2}{16}+\frac{c_2}{4}, \ \ \ p_1=\frac{c_1}{2}.
$$
The coefficients $b_i,e_i,\ i=0,1$ are expressed by $\gamma(n)$,
and $a(n)$ by formulae (6)--(9).
\begin{corollary}
The operator $L(\lambda_1)$ has the following form
$$
 L(\lambda_1)=T^2+u_1(n)T+u_0(n)+u_{-1}(n)T^{-1}+u_{-2}(n)T^{-2},
$$
where
$$
u_1(n)=p_1-e_0(n)-e_0(n+1),
$$
$$
u_0(n)=p_0-b_0(n)-b_0(n+1)-p_1e_0(n)+e_0^2(n)-e_1(n)-e_1(n+1),
$$
$$
u_{-1}(n)=-b_1(n)+b_0(n)\left(-p_1-\frac{b_1(n-1)}{b_0(n-1)}+e_0(n-1)+e_0(n)\right),
$$
$$
u_{-2}(n)=b_0(n)b_0(n-1).
$$
\end{corollary}
The basic result of this work consists of the following:
\begin{theorem}
If we put the function parameters $a(n),\gamma(n)$ equal to
$$
 a(n)=n+1,\ \gamma(n)=n,
$$
then the operators will have coefficients polynomial on $n$. At the
same time,
$$
L_2=T^2+2(n+2)T-\left(\frac{n^4}{2}+n^3-\frac{1}{2}(1-c_2)n^2-\frac{1}{2}(8-c_1-c_2)n\right)-
$$
$$
-\frac{1}{2}(n^3+(c_2-1)n+c_1-2)(n^2+n-1)T^{-1}+
$$
$$
\frac{1}{16}(n^3+(c_2-1)n+c_1-2)(n^3-3n^2+(2+c_2)n+c_1-c_2-2)(n+1)(n-2)T^{-2}.
$$
\end{theorem}

In Theorem 2 we do not list the third order operator due to its
bulkiness. Let's look at an example.
Let a spectral curve be given by the equation
$$
 w^2=z^4+z^2+1,
$$
then
$$
L_2=T^2+2(n+2)T-\frac{1}{2}\left(n^4+2n^3-7n-5\right)-
\frac{1}{2}\left(n^3-2\right)\left(n^2+n-1\right)T^{-1}
$$
$$
+\frac{1}{16}\left(n^3-3n^2+3n-3\right)\left(n+1\right)\left(n-2\right)T^{-2},
$$
$$
L_3=T^3+\left(3n+\frac{15}{2}\right)T^2
-\frac{3}{4}\left(n^4+4n^3+5n^2-8n-14\right)T
$$
$$
-\frac{3}{4}\left(2n^5+7n^4+10n^3+n^2-12n-5\right)
$$
$$
+\frac{3}{16}\left(n^8-2n^6-12n^5-3n^4+10n^3+20n^2+6n-12 \right)T^{-1}
$$
$$
+\frac{3}{32}n\left(2n^2-n-5\right)\left(n^6-3n^5+3n^4-5n^3+6n^2-6n+6\right)T^{-2}-
$$
$$
\frac{1}{64}\left(n-3\right)\left(n^2-1\right)\left(n^3-2\right)\left(n^3-6n^2+12n-10\right)\left(n^3-3n^2+3n-3\right)T^{-3}.
$$

In [7] and [8], we use an analogue of reduction (1) in the smooth
case in order to find commuting differential operators of rank 2,
genus 2.

In section 2 we recall the Krichever-Novikov equation [9] on the discrete Turin parameters dynamics and give formulae for coefficients of operators of rank 2, genus 1 that are expressed by $\zeta$ and $\wp$-Weierstrass functions.

In section 3, theorems 1 and 2 are proven.

\section{Krichever-Novikov equations on the discrete dynamic of the
Turin parameters}

As noted above, for $l>1$ function $\psi(n,P)$ cannot be found in
its explicit form. Let $\Psi(n,P)$ denote a Wronski matrice
$\Psi^{ij}(n,P)=\psi^j(n+i),$ where $\psi^j(n,P)$, $1\leq  j\leq
l$ is some basis in the space of common eigen-functions. As shown
in [9], the number of zeros of functions ${\rm det}\Psi(n,P)$
equals $lg$, where $g$ is the genus of $\Gamma$. Denote them by
$\gamma_1(n),\dots,\gamma_{lg}(n)$. By $\alpha_j(n)$ we denote
vectors where
$$
 \alpha_j(n)\Psi(n,\gamma_j(n))=0.
$$
These vectors are defined up to proportionality. The set
$(\gamma_j(n),\alpha_j(n))$ is called Turin parameters.  The Turin
parameters uniquely define a stable holomorphic bundle of rank $l$
on the curve $\Gamma$. The following theorem, proven in [9],
defines the discrete dynamics of Turin parameters.

\begin{itemize}
\item
The matrix function $\chi(n,P)$ has simple poles in points
$\gamma_j(n)$. The relations on the residues of the matrix
elements
$$
 \alpha_s^j{\rm Res}_{\gamma_s(n)}\chi^{mi}(n,P)=\alpha^i_s(n){\rm
 Res}_{\gamma_s(n)}\chi^{mj}(n,P)  \eqno{(3)}
$$
do hold.

The points $\gamma_s(n+1)$ are zeros of the matrix
determinant $\chi(n,P)$, e.g.
$$
 {\rm det}\chi(n,\gamma_s(n+1))=0.  \eqno{(4)}
$$
The vector $\alpha_j(n+1)$ satisfies the equation
$$
 \alpha_j(n+1)\chi(n,\gamma_j(n+1))=0.  \eqno{(5)}
$$
\end{itemize}

Except for the poles $\gamma_j(n)$ one of the components $\chi$
has a simple pole in the preferred point $Q$.

As follows from the Riemann-Roch theorem, $\chi(n,P)$ is uniquely
restored by the Turin parame\-ters $(\gamma_j(n),\alpha_j(n))$ and
by $l$ arbitrary functional parame\-ters.

This theorem enables us to find operators of rank 2 corresponding to an elliptic curve. Below, we give the according theorem, for
simplicity only for the case where a holomorphic involution on an elliptic
curve switches the location of the poles of function $\chi_1$.
We give the elliptic curve $\Gamma$ as a factor ${\mathbb C}/\Lambda$,
where $\Lambda$ is some lattice in ${\mathbb C}$ and let $z$ be a
coordinate in  ${\mathbb C}$.
Assume that points $\gamma_1 (n)$ and $\gamma_2 (n)$ are interchanged
by a holomorphic involution on $\Gamma$
$$
 \sigma (\gamma_1(n))=\gamma_2 (n),
$$
where $\sigma(z)=-z$. The following theorem was proven in [9].

\begin{itemize}
\item
The operator corresponding to function $\wp (z)$ has the form
$$
 L=L_2^2-\wp (\gamma(n))-\wp (\gamma(n-1)),
$$
where
$$
 L_2=T+v(n)+c(n)T^{-1},
$$
$$
c(n+1)=\frac{1}{4}(s^2(n)-1)F(\gamma(n+1),\gamma(n))F(\gamma(n-1),\gamma(n)),
$$
$$
v(n+1)=\frac{1}{2}(s(n)F(\gamma(n+1),\gamma(n))-s(n+1)F(\gamma(n),\gamma(n+1)),
$$
$$
F(u,v)=\zeta(u+v)-\zeta(u-v)-2\zeta(v),
$$
$s(n),\gamma(n)$ are arbitrary functional parameters.
\end{itemize}
As can be seen from the formulae for the coefficients of operator
$L$, the coefficients are expressed by function
parameters $s(n)$ and  $\gamma(n)$ applying the $\wp$ and
$\zeta$- Weierstrass functions. In corollary 1 and theorem 2 we
found conditions for these coefficients being elementary functions
and, in particular, polynomials.

\section{Proof of theorems 1 and 2}
As follows from our assumption on the invariance $\chi_1(n,P),\
P=(z,w)\in\Gamma$ under the action of involution $\sigma$,
$\chi_1(n,P)$ has the form
$$
 \chi_1(n,P)=\frac{c(n)}{z-\gamma(n)}+d_0(n),
$$
where $c(n),\gamma(n)$ and $d_0(n)$ are some functions of a discrete
variable.
Note that function $\chi_1(n,P)$ has poles in
$$
 P(n)=(\gamma(n),\sqrt{F(\gamma(n)}),\ \ \ \sigma
P(n)=(\gamma(n),-\sqrt{F(\gamma(n)}).
$$
From equalities (4), being equivalent in our case to
$$
 \chi_1(n,P(n+1))=\chi_1(n,\sigma P(n+1))=0
$$
get
$$
\chi_1(n,P)=\frac{c(n)}{z-\gamma(n)}+\frac{c(n)}{\gamma(n)-\gamma(n+1)}.
$$
Look for function $\chi_2$ in the form
$$
 \chi_2(n,P)=\frac{1}{2z}+\frac{a(n)}{2(z-\gamma(n))}+
 \frac{w\gamma(n)}{2z(\gamma(n)-z)}+d(n),
$$
where $a(n)$ and $d(n)$ are some functions.
Note that $\chi_2(n,P)$
has poles in points $P(n), \sigma P(n)$ and $Q$, moreover
${\rm Res}_Q \chi_2=1$.

Vectors $\alpha_1$ and $\alpha_2$ are defined up to proportionality,
therefore we can assume them to have the form
$$
 \alpha_1(n)=(a_1(n),1),\ \alpha_2(n)=(a_2(n),1).
$$
Then from (5) we get
$$
 a_1(n)=-\chi_2(n-1,P(n))=
$$
$$
 \frac{1}{2\gamma(n)}+\frac{a(n-1)}{2(\gamma(n)-\gamma(n-1))}+
 \frac{\sqrt{F(\gamma(n))}
 \gamma(n-1)}{2\gamma(n)(\gamma(n-1)-\gamma(n))}+d(n-1),
$$
$$
 a_2(n)=-\chi_2(n-1,\sigma P(n))=
$$
$$
 \frac{1}{2\gamma(n)}+\frac{a(n-1)}{2(\gamma(n)-\gamma(n-1))}-
 \frac{\sqrt{F(\gamma(n))}
 \gamma(n-1)}{2\gamma(n)(\gamma(n-1)-\gamma(n))}+d(n-1).
$$
Further, the residues of functions $\chi_1$ and $\chi_2$ in poles
$P(n)$ and  $\sigma P(n)$ are equal to
$$
 {\rm Res}_{P(n)}\chi_1={\rm Res}_{\sigma P(n)}\chi_1=c(n),
$$
$$
 {\rm Res}_{P(n)}\chi_2=\frac{1}{2}(a(n)-\sqrt{F(\gamma(n)}),
$$
$$
 {\rm  Res}_{\sigma P(n)}\chi_2=\frac{1}{2}(a(n)+\sqrt{F(\gamma(n)}).
$$
From equalities (3), in our case taking the form
$$
 {\rm Res}_{P(n)}\chi_1=a(n){\rm Res}_{P(n)}\chi_2,
$$
$$
 {\rm Res}_{\sigma P(n)}\chi_1=a(n){\rm Res}_{\sigma P(n)}\chi_2,
$$
get
$$
c(n)=\frac{\gamma(n-1)(a^2(n)-F(\gamma(n))}{4\gamma(n)(\gamma(n)-\gamma(n-1)},
$$
$$
d(n)=\frac{(a(n+1)-1)\gamma(n)+(a(n)+1)\gamma(n+1)}{2(\gamma(n)-\gamma(n+1))\gamma(n+1)}.
$$
Thus, theorem 1 is proven.

Now find the coefficients of operator $L_2$ as follows.
Express $\psi(n+2,P)$ and $\psi(n-2,P)$ by $\psi(n-1,P)$, $\psi(n,P)$,
$\chi_1$ and $\chi_2$.
For this use the identity
$$
 \psi(n+1)=\psi(n-1)\chi_1(n)+\psi(n)\chi_2(n)
$$
from which
$$
\psi(n+2)=\psi(n-1)\chi_1(n)\chi_2(n+1)+\psi(n)(\chi_1(n+1)+\chi_2(n)\chi_2(n+1)),
$$
$$
\psi(n-2)=-\psi(n-1)\frac{\chi_2(n-1)}{\chi_1(n-1)}+\frac{\psi(n)}{\chi_1(n-1)}.
$$
Now replace $T^2\psi(n)$ and $T^{-2}\psi(n)$ in equation
$$
L_2\psi=(T^2+u_1(n)T+u_0(n)+u_{-1}T^{-1}+u_{-2}T^{-2})\psi(n)=\lambda_1\psi(n)
$$
by corresponding expressions. Get
$$
 P_1(n,P)\psi(n,P)+P_2(n,P)\psi(n-1,P)=\lambda_1\psi(n,P),
$$
where
$$
P_1(n)=\chi_1(n+1)+\chi_2(n+1)\chi_2(n)+u_1(n)\chi_2(n)+u_0(n)+\frac{u_{-2}(n)}{\chi_1(n-1)},
$$
$$
P_2(n)=\chi_2(n+1)\chi_1(n)+u_1(n)\chi_1(n)+u_{-1}(n)-u_{-2}(n)\frac{\chi_2(n-1)}{\chi_1(n-1)}.
$$
In the space of common eigen-functions of operators $L_2$ and $L_3$ we
can chose a basis $\psi_1$ and $\psi_2$ normalized by conditions
$$
 \psi_1(n_0,P)=1,\ \psi_2(n_0,P)=0,
$$
with functions $\chi_1,\chi_2$ not depending on normalization points
$n_0$ (see [9]), consequently, we have the identities
$$
 P_1(n,P)-\lambda_1=0,\ \ \  P_2(n,P)=0.
$$
Now, having found in theorem 1 the functions $\chi_1$ and $\chi_2$,
find the coefficients $u_i(n)$ of operator $L_2$ from the series
decomposition $P_1-\lambda_1$ and $P_2$ in the neighbourhood of point $Q$.

The series decomposition coefficients $b_j,e_j$ of functions $\chi_1$
and $\chi_2$ in the neighbourhood of $Q$ are expressed by
the functional parameters $a(n)$ and $\gamma(n)$ according to
formulae
$$
b_0(n)=\frac{\gamma(n-1)\gamma(n+1)(F(\gamma(n))-a^2(n))}{4(\gamma(n-1)-\gamma(n))(\gamma(n)-\gamma(n+1))\gamma(n)^2},
 \eqno{(6)}
$$
$$
b_1(n)=\frac{\gamma(n-1)(a^2(n)-F(\gamma(n))}{4(\gamma(n-1)-\gamma(n))\gamma^3(n)},\eqno{(7)}
$$
$$
e_0(n)=\frac{1}{2}\left(\frac{c_1}{2}+\frac{1}{\gamma(n)}-\frac{a(n)}{\gamma(n)}+
 \right.
$$
$$
\left.\frac{(a(n+1)-1)\gamma(n)+(a(n)+1)\gamma(n+1)}{(\gamma(n)-\gamma(n+1))\gamma(n+1)}\right),\eqno{(8)}
$$
$$
e_1(n)=\frac{8-8a(n)+4c_1\gamma(n)-(c_1^2-4c_2)\gamma^2(n)}{16\gamma^2(n)}.\eqno{(9)}
$$
Analogically, find the operator corresponding to a meromorphic function
with one third order pole in $Q$.
Direct verification yields that for
$$
 a(n)=n+1,\ \gamma(n)=n
$$
the operators' coefficients are polynomials on $n$. Thus, theorem 2 is
proven.

\vskip7mm {\bf References}

\vskip3mm

[1] Burchnall, J.L.; Chaundy, I.W. Commutative ordinary
differential operators // Proc. London Math. Society. 1923. Ser.
2. V. 21. P. 420--440.

[2] Krichever, I.M. Commutative rings of ordinary differential
operators
// Funk. Anal. Pril. 1978. V. 12. N. 3. P. 20--31.

[3] Dixmier, J. On Wiyl algebras // Matematika. 1969. V.13, N. 4.
P. 16--44.

[4] Krichever, I.M.; Novikov S.P. Holomorphic vector bundles on
Riemann surfaces and the Kadomtcev--Petviashvili equation
// Funk. Anal. Pril. 1978. V. 12. N. 4. P. 41--52.

[5] Grinevich, P.G. Rational solutions of the commutativity
equation of differential operators.
// Funk. Anal. Pril. 1982. V. 16. N. 1. P. 19--24.

[6] Mokhov, O.I. Commuting differential operators of rank 3, and
nonlinear equations// Izv. Akad. Nauk SSSR Ser. Matem. 1989. V.
53. N. 6, P. 1291--1314.

[7] Mironov, A.E. A ring of commuting differential operators of
rank 2 corresponding to a curve of genus 2 // Sbornik Mathematics.
 2004. V. 195. N. 5. P. 711-722.

[8] Mironov, A.E. Commuting Rank 2 differential operators
corresponding to a curve of genus 2 // Funct. Anal. and its Appl.
2005. V. 39. N. 3. P. 240--243.

[9]  Krichever, I. M., Novikov, S. P. A two-dimensionalized Toda
chain, commuting difference operators, and holomorphic vector
bundles // Russ. Math. Surveys. 2003. V. 58, N. 3, P. 473--510

[10] Mumford, D. An algebro-geometric construction of commuting
opera\-tors and  of solution to the Toda lattice equation,
Kortweg-de Vries equation and related nonlinear equations //
Proceedings of the International Symposi\-um on Algebraic Geometry
(Kyoto Univ., Kyoto, 1977). Tokyo: Kinokuniya Book Store, 1978. P.
115--153.

[11] Krichever, I. M. Algebraic curves and nonlinear difference
equations // Uspekhi Matem. Nauk. 1978. V. 33, N. 4, P. 215--216.
$$
$$

Sobelev Institute of Mathematics, 630090 Novosibirsk, Russia;

E-mail: mironov@math.nsc.ru

\end{document}